\documentclass[11pt,twoside]{article}
\usepackage{jltp}
\usepackage{psfig}

\title{Pseudogap and Superconducting Gap in YBa$_2$Cu$_3$O$_{6+x}$:
A~Raman~Study}

\author{Matthias~Opel, Martin~G\"otzinger, Christian~Hoffmann, 
Ralf~Nemetschek, Richard~Philipp, Francesca~Venturini,
Rudi~Hackl\address{Walther--Meissner--Institut, Bay. Akad. d. Wiss.,
D--85748 Garching, Germany},
Andreas~Erb$^\ast$, and Eric~Walker\address{DPMC,
Universit\'e de Gen\`eve, CH-1211 Gen\`eve, Switzerland}}

\runninghead{M. Opel {\it et al.}}
{Pseudogap and s.c. Gap in Y--123: A Raman Study}
       
\begin{document}

\begin{abstract}
We present results of electronic Raman-scattering experiments
in differently doped YBa$_2$Cu$_3 $O$_{6+x}$.
In B$_{2g}$ symmetry, an analysis of the data in terms 
of a memory function approach is presented and
dynamical relaxation rates $\Gamma(\omega,T)$
and mass-enhancement factors $1+\lambda(\omega,T)$
for the carriers are obtained. 
Starting from temperatures $T > 180 {\rm K}$, 
$\Gamma(\omega,T)$ and $1+\lambda(\omega,T)$ are extrapolated
to lower temperatures and used to re-calculate Raman spectra.
By comparison with our data, we find
a loss of spectral weight between $T_c < T < T^\ast$
at all doping levels $x$.
$T^\ast$ is comparable to the pseudogap temperature
found in other experiments. 
Below $T_c$, the superconducting gap is observed.
It depends on $x$ and scales with $T_c$
whereas the energy scale of the pseudogap remains the same.

\end{abstract}

\maketitle

%Include this space if you do not use sections in your document.
%\vspace{0.3in}

\section{INTRODUCTION AND EXPERIMENTAL}

The relationship between the superconducting (SC) and the pseudogap (PG) phases
and their evolution with doping is of particular interest for understanding
the cuprates. In the following we will describe
recent results from light scattering experiments in differently doped,
high--quality YBa$_2$Cu$_3$O$_{6+x}$ (Y--123) single crystals \cite{purity}.
The experiments were performed in pseudo back-scattering geometry
using a standard Raman set-up.
We present an analysis of the spectra in terms of a memory function approach
which is well established for optical and infrared (IR)
spectroscopy \cite{timusk99}.
We will focus on B$_{2g}$ symmetry only
since it represents carrier properties independently of the 
doping level \cite{JLTP}.

\section{MODEL AND ANALYSIS}

\begin{figure}[htb]
\centerline{\psfig{file=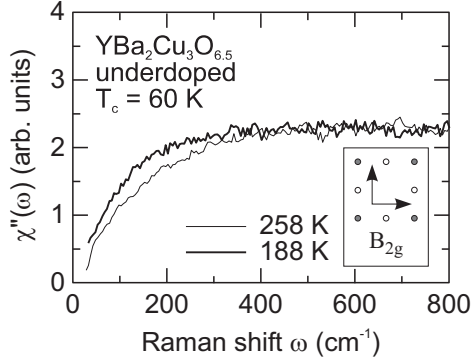}}
\caption{Raman response $\chi''(\omega,T)$
for underdoped Y--123.}  
\label{fig:response}
\end{figure}

\begin{figure}[htb]
\centerline{\psfig{file=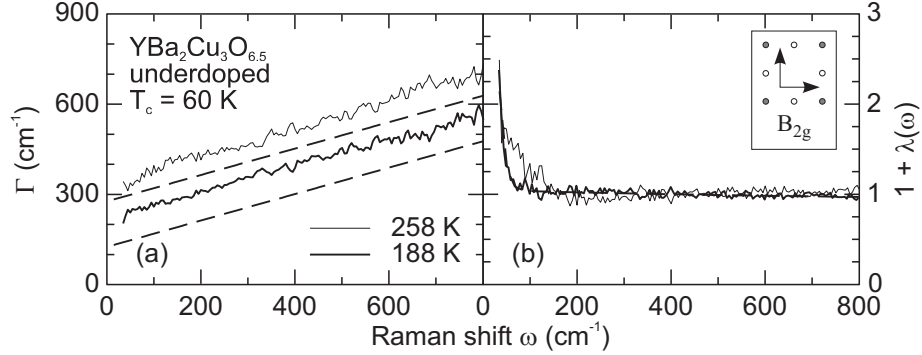}}
\caption{(a) Dynamical relaxation rate $\Gamma(\omega,T)$ and
(b) mass renormalization $1+\lambda(\omega,T)$
for the spectra shown in Fig.~\ref{fig:response}.
The dashed lines are extrapolations for 230K and 70K, respectively.}
\label{fig:GammaLambda}
\end{figure}

For the study of the dynamical response we adopt the approach 
in terms of a memory function 
\begin{equation}
M(\omega,T) = i\Gamma(\omega,T) + \omega\lambda(\omega,T)
\label{eq:MemoryFunction}
\end{equation}
with the carrier relaxation rate $\Gamma=1/\tau$
and the mass enhancement factor $1+\lambda=m^\ast/m$.
The formalism was introduced by G\"otze and W\"olfle for the  
current-current correlation function \cite{memory} and subsequently 
applied to the analysis of IR data \cite{timusk99}. 
In our Raman experiment, 
the number of inelastically scattered photons 
registered per unit time is proportional to 
the imaginary part $\chi''$ of the response function  
$\chi = R M/(\omega+M)$
which reads
\begin{equation}  
\chi''(\omega,T) = R \frac{\omega\Gamma(\omega,T)}
{\omega^2(1+\lambda(\omega,T))^2+\Gamma^2(\omega,T)} 
\label{eq:chi"}  
\end{equation}  
with $R$ a symmetry-dependent scale factor.
Defining $I=\chi''/\omega$, we obtain the following expressions 
\begin{eqnarray}  
\Gamma(\omega,T) & = & R \frac{I(\omega,T)}
{I^2(\omega,T) + \omega^2 K^2(\omega,T)}
\label{eq:Gamma}
\\
1+\lambda(\omega,T) & = & R \frac{K(\omega,T)}
{I^2(\omega,T) + \omega^2 K^2(\omega,T)}
\label{eq:lambda}  
\end{eqnarray}  
where $K(\omega,T)$ is given by
\begin{equation}  
K(\omega,T) = - \frac{2}{\pi}  
\wp\int_{0}^{\infty} d\xi \frac{I(\xi,T)}{\xi^2-\omega^2}.  
\label{K-I}
\end{equation}  
Applying this model to the Raman spectra from underdoped Y--123
in the normal state (Fig.~\ref{fig:response}) 
we find $\Gamma$ to depend linearly on $\omega$
and $1+\lambda$
to be almost $T$--independent and close to unity for a 
fairly large frequency range (Fig.~\ref{fig:GammaLambda}).
A more detailed discussion of this formalism 
will be presented in a forthcoming publication.

\begin{figure}[htb]
\centerline{\psfig{file=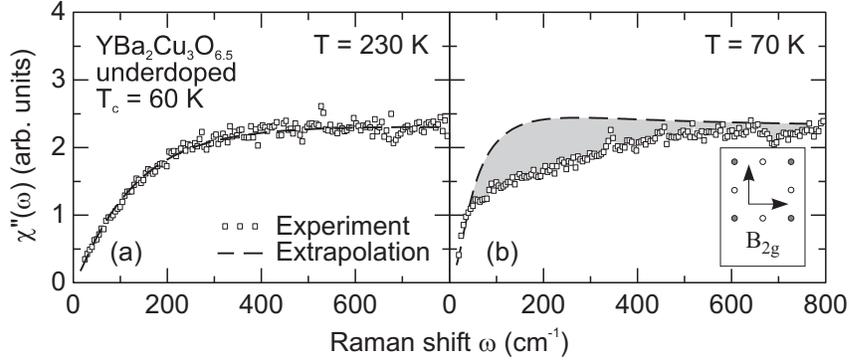}}
\caption{Comparison between the measured $\chi''_{\rm exp}$ (squares)
and the extrapolated Raman response $\chi''_{\rm ext}$ (dashed lines) 
for (a)~230K and (b)~70K.}
\label{fig:ModelCheck}
\end{figure}

The $\omega$--dependence of $\Gamma$ and $1+\lambda$ 
can be fitted by monotonous functions
\begin{eqnarray}  
\Gamma(\omega,T) &=& \Gamma_0(T) + \alpha\omega
%\Gamma(\omega,T) &=& \Gamma_0(T) + 0.44\omega
\label{eq:FitGamma}   
\\
1+\lambda(\omega) &=& 1+ \lambda_0 
+ \lambda_1 e^{-\omega/\omega_1} + \lambda_2 e^{-\omega/\omega_2}.
%1+\lambda(\omega) &=& 0.853+32e^{-\omega/10.5{\rm cm}^{-1}} + 0.14e^{-\omega/945{\rm cm}^{-1}}
\label{eq:FitLambda}  
\end{eqnarray}
with fitting parameters $\alpha$, $\lambda_{0,1,2}$, 
and $\omega_{1,2}$. This allows us to extrapolate Raman spectra 
measured at $T$ to different temperatures $T'$.
In the following, this will be done for $T'=230$K and $T'=70$K.
We simply put $\Gamma_0(T')$ equal to the static limit of the
relaxation rate $\Gamma(0,T')$ which can be determined directly 
from the spectra at $T'$ while leaving $\alpha$, $\lambda_{0,1,2}$, and
$\omega_{1,2}$ unchanged and obtain $\Gamma$ and
$1+\lambda$ via Eqs.~\ref{eq:FitGamma} and~\ref{eq:FitLambda}
(dashed lines in Fig.~\ref{fig:GammaLambda}).
Using Eq.~\ref{eq:chi"} we now can calculate "extrapolated" spectra
$\chi''_{\rm ext}$. By comparing the 230K extrapolation
to our experimental data $\chi''_{\rm exp}$
we see that this procedure is fairly reliable
(Fig.~\ref{fig:ModelCheck}~(a)). However, at 70K we find 
a deviation $\Delta\chi''=\chi''_{\rm exp}-\chi''_{\rm ext}$
(shaded area in Fig.~\ref{fig:ModelCheck}~(b))
which has been attributed to the opening of a PG \cite{RamanPG}. 
Now the presented extrapolation procedure
allows us to quantify $\Delta\chi''$ 
as a function of temperature and doping.

\section{RESULTS AND DISCUSSION}

\begin{figure}[htb]
\centerline{\psfig{file=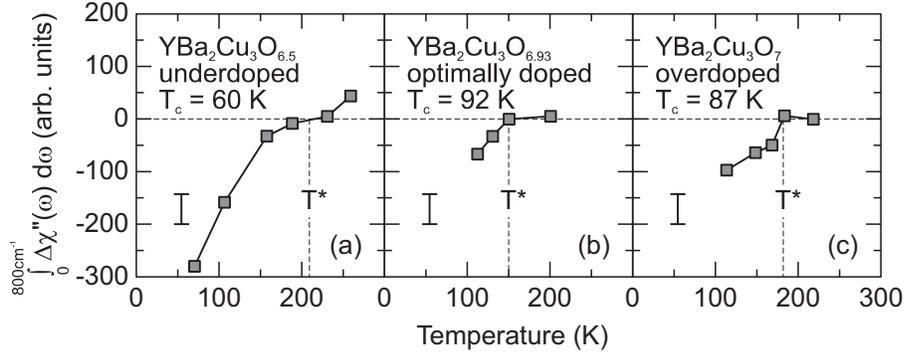}}
\caption{Integrated change of spectral weight $\Delta\chi''$ 
in the normal state as a function of temperature for 
(a)~underdoped, (b)~optimally doped, and (c)~overdoped Y--123.
The vertical bars represent the experimental errors.}
\label{fig:SpectralWeight}
\end{figure}

Figure~\ref{fig:SpectralWeight} shows $\Delta\chi''$,
integrated from $\omega=0$ to $\omega=800{\rm cm}^{-1}$ 
as a function of temperature.
In all doping levels,
spectral weight is lost below a characteristic 
temperature~$T^\ast$ (vertical dashed lines in
Fig.~\ref{fig:SpectralWeight}) which compares well 
with the pseudogap temperature found in other 
experiments \cite{timusk99}.
This effect reaches 17\% in the underdoped sample
and becomes much weaker towards optimal doping.

\begin{figure}[htb]
\centerline{\psfig{file=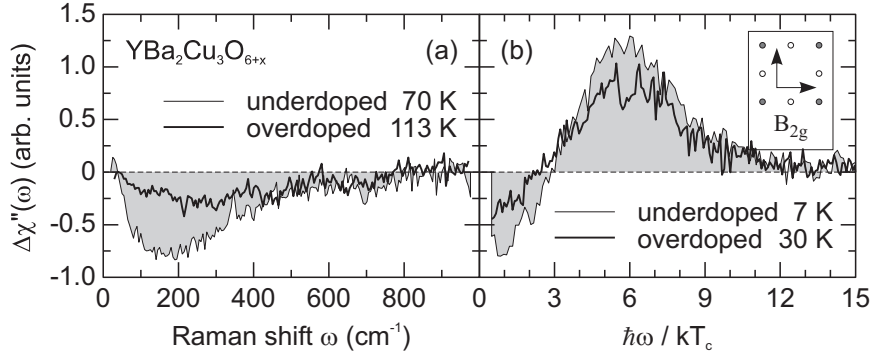}}
\caption{Change of spectral weight $\Delta\chi''$ 
(a)~in the PG state and (b)~in the SC state for 
underdoped ($T_c=60$K) and overdoped Y--123 ($T_c=87$K).}
\label{fig:PGandSC}
\end{figure}

To study the influence of doping on the energy range
of the PG $\Delta^\ast$ we choose the lowest temperatures
(where the effect becomes largest) in 
Figs.~\ref{fig:SpectralWeight}~(a) and~(c) and plot
$\Delta\chi''(\omega,T)$ as a function of $\omega$.
Fig.~\ref{fig:PGandSC}~(a) shows that the presence of the PG
leads to a suppression of spectral weight up to 800cm$^{-1}$
in both cases. The energy scale of $\Delta^\ast$ seems to be
fixed and independent of the doping level. 
This is not the case for the $\omega$-range of the SC correlations 
which clearly scales with the SC transition temperature $T_c$
(Fig.~\ref{fig:PGandSC}~(b)). The breaking of Cooper pairs leads 
to an increase of spectral weight between 3 and $12kT_c$
for both doping levels. This means that the absolute value of
the SC energy gap $\Delta$ has to increase by 45\%
when going from the underdoped ($T_c=60{\rm K}$) to the
overdoped ($T_c=87{\rm K}$) regime. A similar behavior has
been found for 
Bi$_2$Sr$_2$CaCu$_2$O$_{8+\delta}$ \cite{JLTP} and
La$_{2-x}$Sr$_x$CuO$_4$ \cite{Naeini}.

In summary, our Raman experiment allows to compare $\Delta$
and $\Delta^\ast$ in the same sample. We find that
whereas the range of finite $\Delta$ scales with $T_c$,
the energy range of $\Delta^\ast$ is independent of doping.
As a result, our experiments do not support evidence
for a relation between the PG and the SC gap.

\section*{ACKNOWLEDGMENTS}
We gratefully acknowledge continuous support by 
B.S.~Chandrasekhar, D.~Einzel, and I.~T\"utt\H{o}.
We are indebted to N.~N\"ucker  
who cut some of the crystals with the microtome.
We are grateful to the BMBF for financial support 
via the program ``Bilaterale 
Wissen\-schaft\-lich-Tech\-nische 
Zusammen\-arbeit'' under grant no. UNG-052-96.

\end{document}